\documentclass[preprint,amsmath,aps]{revtex4}

\usepackage{graphicx}
\usepackage{dcolumn}
\usepackage{bm}

\begin{document}


\title{Quantum states and localisation of developable M\"obius nanostructures}

\author{A.~P.~Korte}
\author{~G.~H.~M.~van der Heijden}
\affiliation{
Centre for Nonlinear Dynamics, University College London, Chadwick Building, Gower Street, London WC1E 6BT, UK
}

\date{\today}

\begin{abstract}
The equilibrium equations for wide, developable, M\"obius strips undergoing large deformations have recently been derived, and solved, numerically. We use these results to compute the eigenvalues and eigenstates of non-interacting electrons confined to M\"obius strips of linking number up to $1.5$ and of arbitrary width. The inverse participation ratio is used to show that electrons are increasingly localised to the higher curvature regions of the higher-width structures, where sharp creases could form channels for particle transport. Our geometric formulation could be used to study transport properties of M\"obius strip and other components in nanoscale devices.
\end{abstract}


\maketitle

\section{\label{sec:level1}Introduction}

The eigenstates of a particle confined to an inextensible M\"obius strip have previously been calculated, assuming the Schwarz parametrisation  \cite{Grave1} as an approximation to the centreline of the strip, for the single, large, aspect ratio, $L/2w=30$, corresponding to the dimensions of ribbon-shaped crystals of NbSe$_3$ M\"obius shell structures that have recently been fabricated \cite{Tanda1}. For this aspect ratio, using the Schwarz parametrisation, the bending energy is easily minimised, after performing the Wunderlich reduction \cite{Gert,Wunderlich} to a one-dimensional integral over the centreline of the strip. The constraint that the length of the strip must equal $L$ allows one of the three Schwarz parameters to be eliminated, giving a two-dimensional visualisation of the bending energy in the remaining two Schwarz parameters. For $L/2w=30$, one minimum is clearly visible. For smaller aspect ratios (larger widths), however, as considered in \cite{Gert}, e.g. $L/2w=1$, there are three candidate minima, none of which satisfy the constraint which ensures the generators do not intersect within the strip \cite{Grave1,Wunderlich}, $|\eta'(u)|\leq w^{-1}, u\in [0,2\pi]$ (for $\eta$, see below), indicating the limitations of the Schwarz parametrisation even for modest strip widths. 

We note that recently \cite{Caetano}, wave functions have been calculated for graphene nanoribbons of varying $\mathrm{Lk}$, again for a single, large, value of $L/2w=6.8$. After a classical geometry optimisation for a discrete (atomistic) model, an annealing simulation searches for lowest energy structures of a classical dynamics. The electronic structure is simulated using a semi-empirical Hamiltonian.

In order to enable, and generalise, the calculation of energy eigenvalues and surface eigenstates of a non-relativistic quantum particle to arbitrary widths, therefore, we use the formulation in \cite{Gert} which rigorously minimises the Wunderlich elastic bending energy functional \cite{Wunderlich} to obtain the exact shape of M\"obius strips of linking numbers $\mathrm{Lk}=0.5, 1.5$. Consideration of the large width behaviour of the inverse participation ratio provides evidence for curvature trapping in these periodic structures normally seen in infinite domain \cite{Hurt} or disordered systems \cite{Thouless}.

\section{The quantum mechanics of a particle bound to a surface}
In the formulation \cite{Costa} for an infinitely strong squeezing potential which attaches a particle to a surface, the transformation of the wave function, $\psi$, to $\chi=\sqrt{F}\psi$, with $F=1-2Mu^2+K(u^3)^2$, leads to a separation of the three-dimensional Schr\"odinger equation for $\chi=\chi_t(u^1,u^2)\chi_n(u^3)$, where $u^1, u^2$ are the coordinates embedded in the surface, and $u^3$ is the distance perpendicular to it, giving for the surface wave function $\chi_t$:
\begin{equation}\label{schr_eq}
-\frac{\hbar^2}{2m}(\Delta+M^2)\chi_t=E\chi_t,
\end{equation}
where $\Delta$ is the Laplace-Beltrami operator on the surface.
For a developable (inextensible) surface, the Gaussian curvature $K$ is zero, and the mean curvature is easily calculated, using the coefficients of the first and second fundamental forms of the surface, as
\begin{equation}
M=-\frac{\kappa}{2}\frac{1+\eta^2}{1+t\eta'}.
\end{equation}
If $\mathbf{r}(s)$ is a parametrisation of the centreline of the strip, then 
\begin{equation}
\begin{split}
\mathbf{x}(s,t)=\mathbf{r}(s)+t\left[\mathbf{b}(s)+\eta(s)\mathbf{t}(s)\right]\\
\tau(s)=\eta(s)\kappa(s), \quad s\in [0,L], t\in[-w,w]
\end{split}
\end{equation}
is a parametrisation of a strip of length $L$ and width $2w$, where $\mathbf{t}$  is the unit tangent vector, $\mathbf{b}$ the unit binormal and where $\kappa$ and $\tau$ are respectively the curvature and torsion of the centreline \cite{Randrup} which uniquely specify (up to Euclidean motions) the centreline of the strip as well as the Serret-Frenet basis vectors $\mathbf{t}, \mathbf{n}, \mathbf{b}$. 
The surface is developable and is completely determined by the centreline of the structure. Developing the surface into a rectangle with rectangular coordinates $(u^1,u^2)$, given by 
\begin{equation}\label{u1_u2}
u^1=s+t\eta(s), \quad u^2=t,
\end{equation}
$\Delta$ in \eqref{schr_eq} is then the usual Laplacian and $M^2$ provides a quantum potential well.
The boundary conditions are
\begin{equation}\label{bcgen}
\begin{split}
\chi(u^1,u^2=-w)&=\chi(u^1,u^2=w)=0,\\
\chi(u^1+2L,u^2)&=\chi(u^1,u^2),
\end{split}
\end{equation}
where the latter is the requirement of the single-valuedness of the wave function \cite{single_val2}, \cite{single_val1}, known as the periodic (or Born-von Karman) boundary condition \cite{Mermin}.

\begin{figure}[floatfix]
\includegraphics[width=0.9\textwidth]{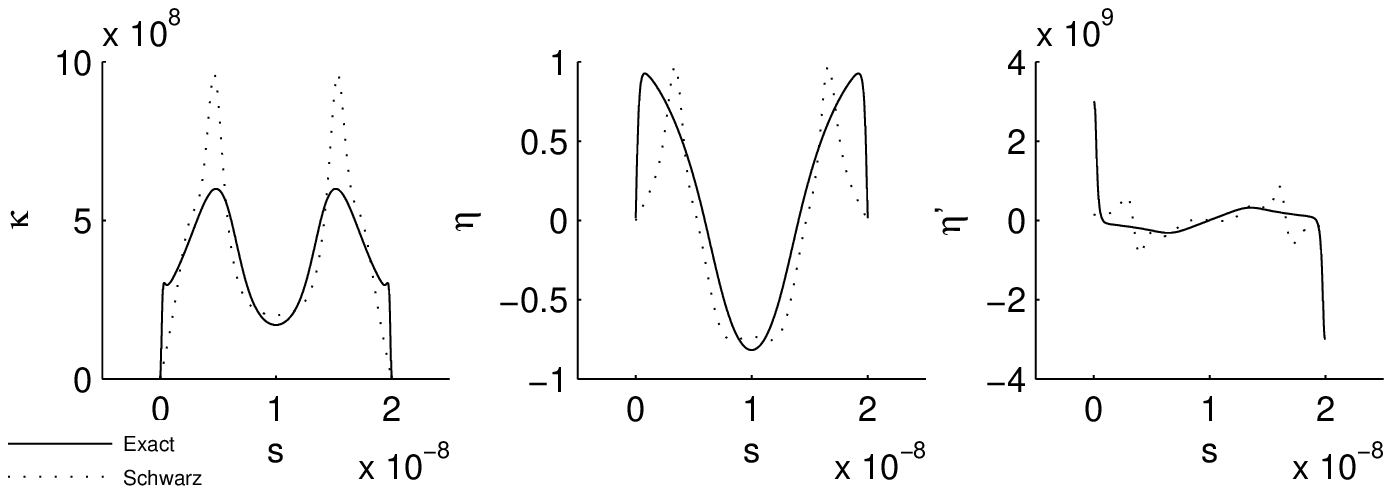}
\caption{\label{fig:kappa}$\mathrm{Lk}=0.5, L/2w=30$}
\end{figure}

\section{Numerical Results and discussion}
\subsection{Single twist M\"obius strip: $\mathrm{Lk}=0.5$}

Shown in Fig.~\ref{fig:kappa} are $\kappa, \eta, \eta'$ as a function of the arc-length, $s$, in \cite{Gert} for the exact shape of a free-standing M\"obius strip, characterised solely by its bending energy, for $L/2w=30$, compared with the Schwarz parametrisation.
Clearly $\kappa$ and $\eta$ have an even reflection symmetry about $s=L/2$ whereas $\eta'$ has odd reflection symmetry which through \eqref{u1_u2} induce the following transformation in $(u^1,u^2)$:
\begin{equation}\label{refl_transf}
u^1\rightarrow L-u^1, \qquad u^2\rightarrow-u^2.
\end{equation}
The Hamiltonian $\mathcal{H}=-\frac{\hbar^2}{2m}(\nabla^2+M^2)$ is invariant under this transformation since $M^2$  is. Therefore non-degenerate states are such that 
\begin{equation}\label{chi_refl}
\chi(u^1,u^2)=\pm\chi(L-u^1,-u^2).
\end{equation}
Similarly $\mathcal{H}$ is invariant under the transformation
\begin{equation}\label{transl_transf}
u^1\rightarrow u^1+L, \qquad u^2\rightarrow-u^2,
\end{equation}
itself induced by the transformation $s\rightarrow s+L$, with corresponding parity eigenstates
\begin{equation}\label{chi_trans}
\chi(u^1,u^2)=\pm\chi(u^1+L,-u^2).
\end{equation}
Eq.~\eqref{chi_trans} can be recognised as the Bloch or Floquet theorem for periodic potentials:
\begin{equation}\label{Bloch}
\chi(u^1+L,u^2)=\mathrm{e}^{ikL}\chi(u^1,-u^2),
\end{equation}
which, given the periodic boundary condition, gives $kL=n\pi$, with $n$ an integer. Eq.~\eqref{Bloch} then gives \eqref{chi_trans}.
Thus four different symmetry eigenstates are considered: even and odd reflection symmetry in the line $u^1=L/2$, and even and odd symmetry under translation by $L$, each with $u^2\rightarrow -u^2$. Reflection symmetry, \eqref{chi_refl}, allows the domain for the numerical computation to be reduced to half the strip. We used a finite-difference (FD) scheme, calculating the eigenvalues and eigenstates with MATLAB.  Thus in the FD scheme, using \eqref{transl_transf} for $u^1<0$, $M^2(u^1,u^2)=M^2(2L-|u^1|,u^2)=M^2(L-|u^1|,-u^2)$, (as $\mathbf{b}$ has changed sign and we are exactly under the surface from where we started), $=M^2(L+|u^1|,-u^2)$, (by symmetry), $=M^2(|u^1|,u^2)$ (as $\mathbf{b}$ has changed sign). This enables us to distinguish between particles on the strip from those in the strip (cf. \cite{Maiti} for a flat M\"obius lattice model), the former allowing negative parity eigenstates under translation by $L$. Without imposing \eqref{chi_refl}, we find the FD scheme skips some eigenstates. The results were also checked with finite elements. The requirement $u^2\rightarrow -u^2$ for invariance of $M^2$ under the $s\rightarrow s+L$ translation is because we use a continuously varying $\mathbf{t},\mathbf{n},\mathbf{b}$ frame moving along the centreline, changing to an anti-Frenet frame to avoid a Frenet frame flip where $\kappa=0$  ($\mathbf{n}\rightarrow-\mathbf{n}$), which therefore defines the co-ordinate system used in $u^1$ and $u^2$. In addition, we therefore require that $\eta\rightarrow-\eta$ and $\eta'\rightarrow-\eta'$ and $\kappa\rightarrow-\kappa$, under $s\rightarrow s+L$, to define the $\mathbf{t},\mathbf{n},\mathbf{b}$ frame used. Otherwise, if the Frenet frame is used throughout, it flips under translation, and $u^2\rightarrow -u^2$ would not be required.

\begin{figure}[floatfix]
$
\begin{array}{cc}
\includegraphics[width=0.9\textwidth]{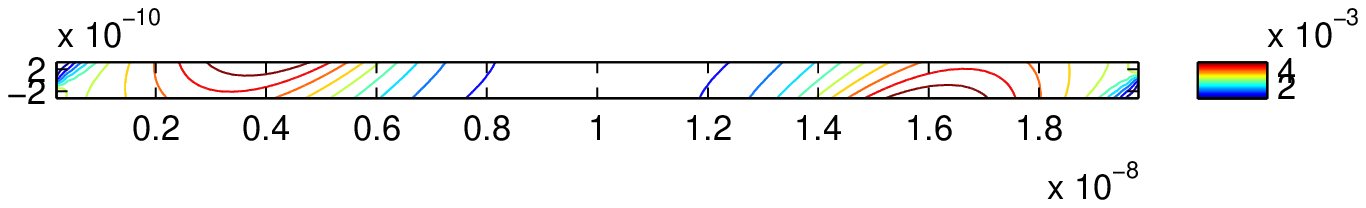} \\
\includegraphics[width=0.9\textwidth]{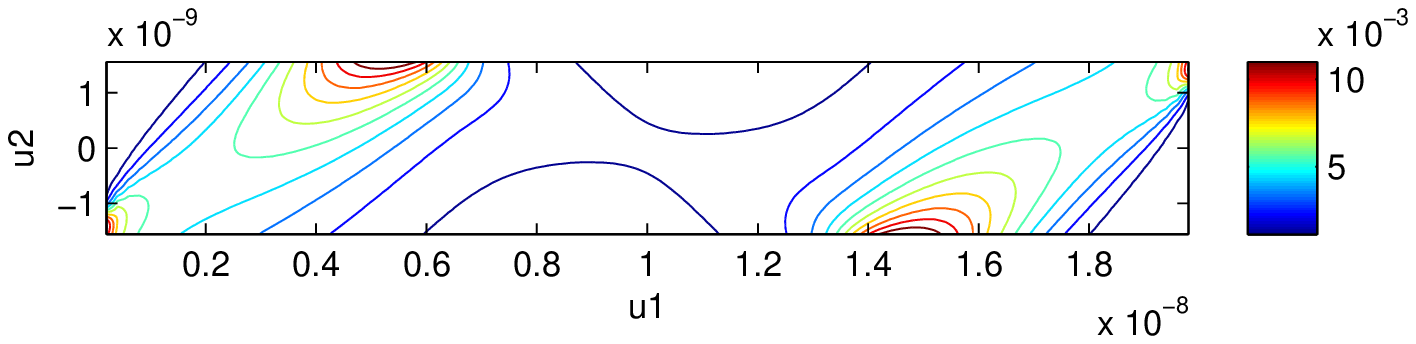} \\
\includegraphics[width=0.9\textwidth]{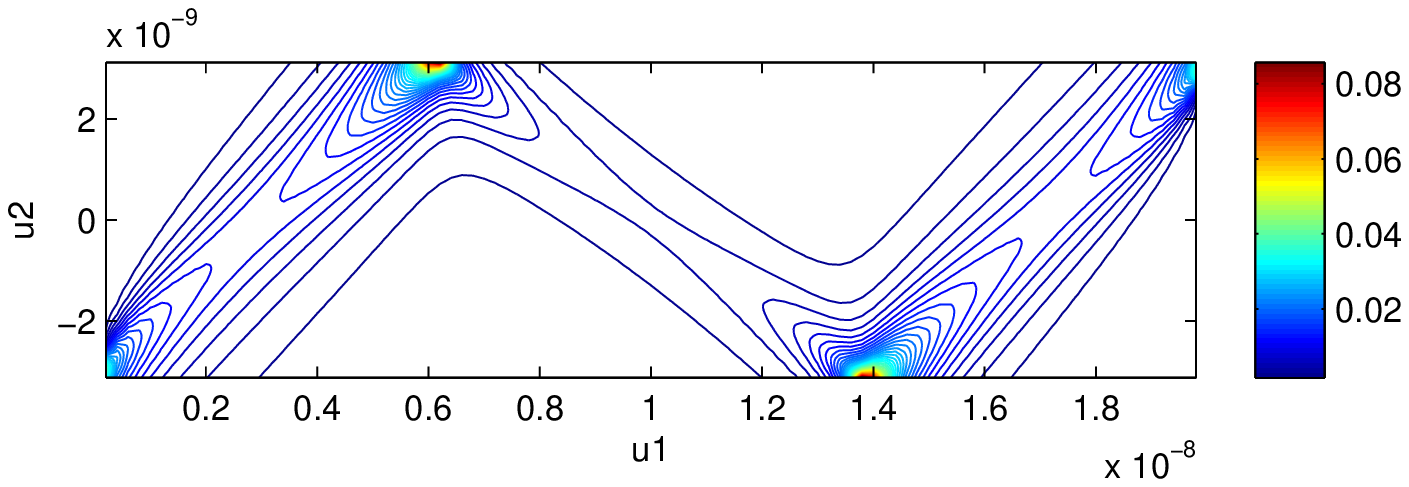} \\
\includegraphics[width=0.9\textwidth]{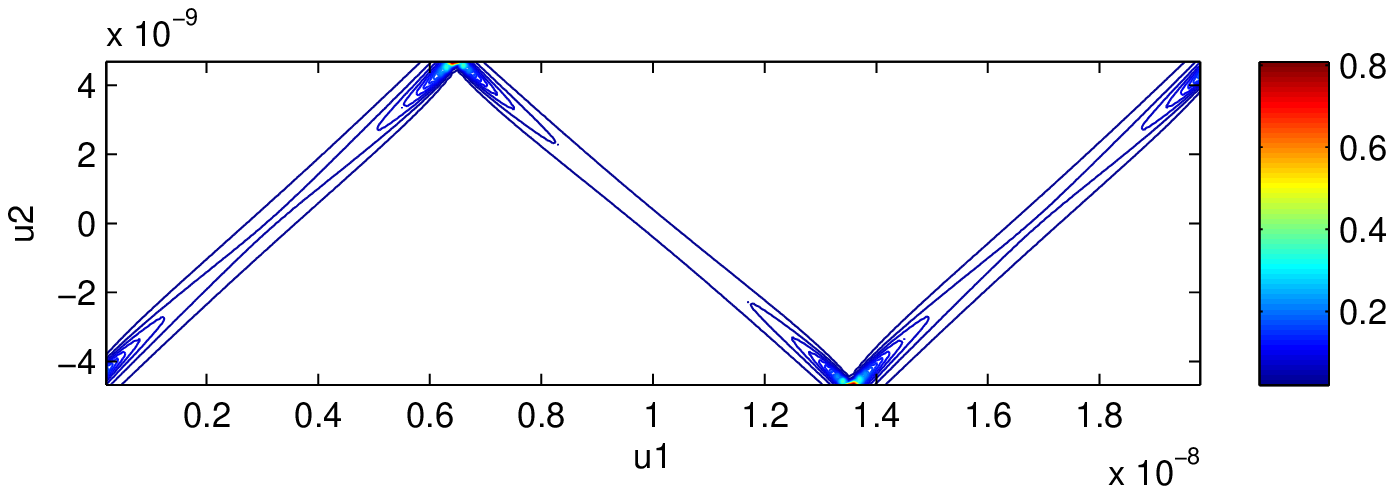}
\end{array}$
\caption{\label{fig:Msq_Lk_0_5}$M^2$ in eV, $\mathrm{Lk}=0.5$, $L/2w=30,2\pi,\pi,2\pi/3$}
\end{figure}

\begin{figure}[floatfix]
\includegraphics[width=0.9\textwidth]{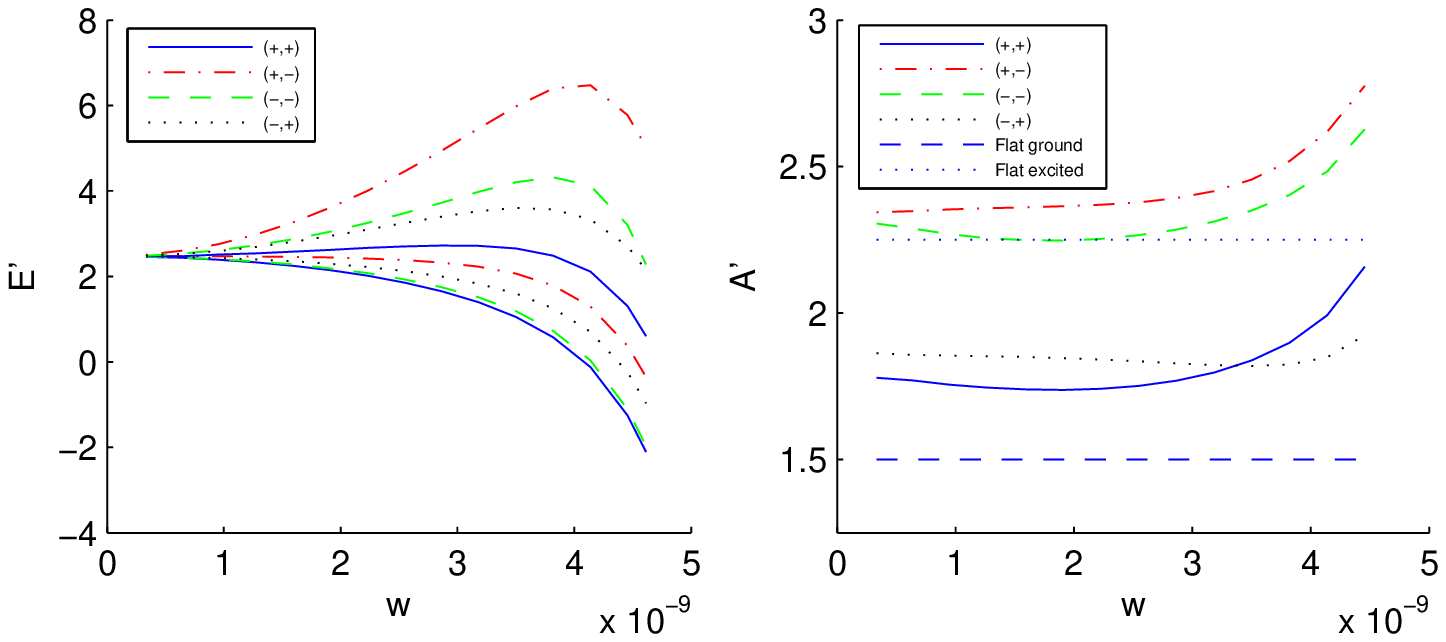}
\caption{\label{fig:AE_Lk_0_5}$E',A'$, $\mathrm{Lk}=0.5$}
\end{figure}

Shown in Fig.~\ref{fig:Msq_Lk_0_5} are contour plots of $M^2$ for four increasing widths, taken from \cite{Gert}, which form the potential wells which scatter the standing waves of electrons confined to the strip. For clarity, the outer boundary is omitted, as there are singularities there for $|\eta'(s)|=1/w$ \cite{Gert}. As $w$ increases, creases are formed in the M\"obius structure, which a quantum particle experiences as deepening potential wells which lower its energy. For low enough aspect ratio ($L/2w=10\pi/13$), negative energy eigenstates appear, the usual signature for bound states.
Shown in Fig.~\ref{fig:AE_Lk_0_5} (left) is the dimensionless energy $E'=(\hbar^2 w^2/2m)E$ versus $w$. By comparison with the non-geometric (flat) strip (obtained with $M^2$ in \eqref{schr_eq} set to zero), we expect the energy to decrease with increasing width. There are therefore two competing effects. Multiplying $E$ by $w^2$ makes $E'$ increase with width initially, but for larger widths, a decreasing $E'$ means that $E$ decreases faster than $w^2$. The figure follows the lowest energy eigenstate for each parity with increasing width, with either one or two nodes in the $u^1$ direction, where, we denote, for example, a state as (-,+), if it is odd under translation by $L$, and even under reflection.

\subsection{Triple twist M\"obius strip: $\mathrm{Lk}=1.5$}

\begin{figure}[floatfix]
\includegraphics[width=0.9\textwidth]{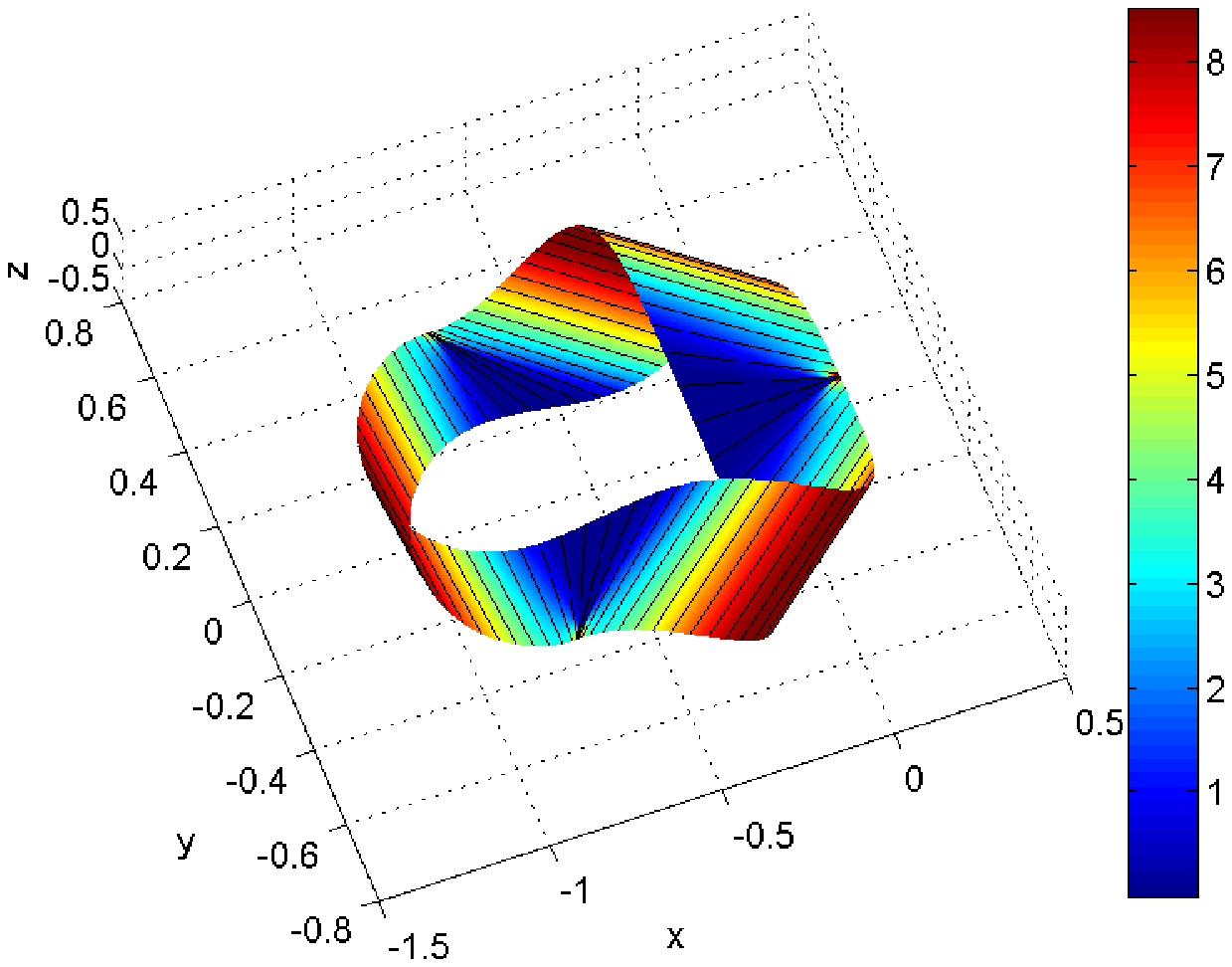}
\caption{\label{fig:surface_Msq_Gens} $\mathrm{Lk}=1.5$, unscaled $L/2w=9.36$ structure with $M^2$ and generators shown.}
\end{figure}

\begin{figure}[floatfix]
$
\begin{array}{cc}
\includegraphics[width=0.9\textwidth]{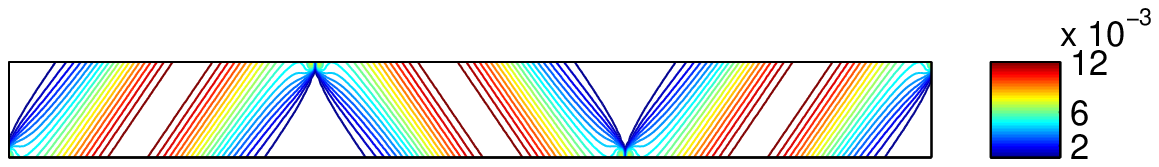} \\
\includegraphics[width=0.9\textwidth]{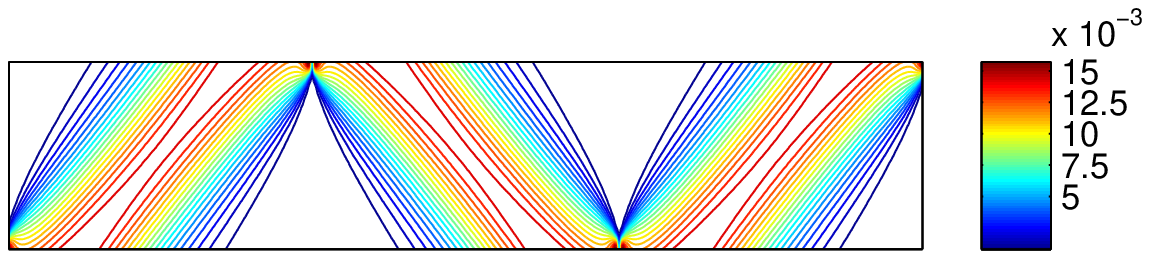} \\
\includegraphics[width=0.9\textwidth]{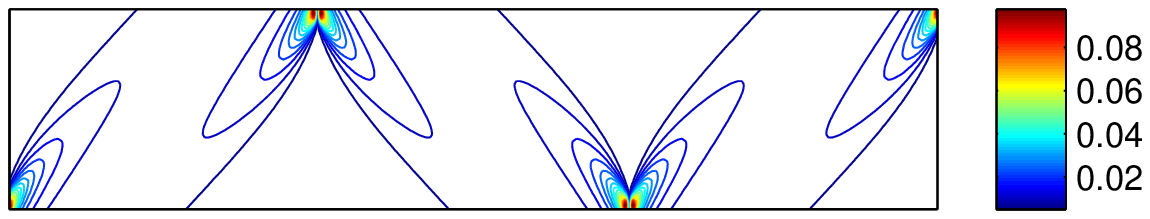} \\
\includegraphics[width=0.9\textwidth]{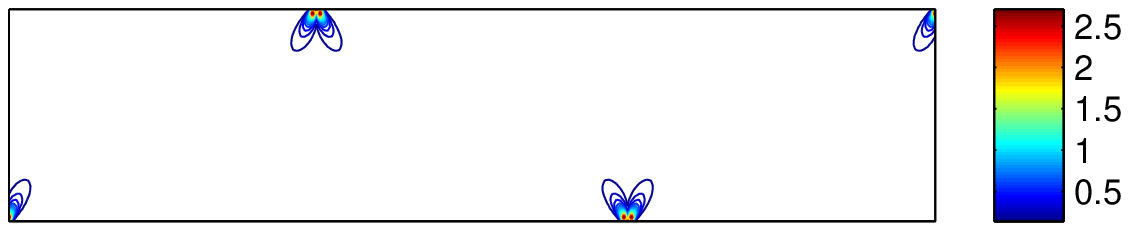}
\end{array}$
\caption{\label{fig:Msq_Lk_1_5} $M^2$ in eV, $\mathrm{Lk}=1.5$, $L/2w=9.36,4.72,4.49,4.22$.}
\end{figure}

Shown in Fig.~\ref{fig:surface_Msq_Gens} is an unscaled triple twist M\"obius structure, constructed solely from the values of $\kappa$ and $\tau$ of its centreline in \cite{Gert}. The corresponding contour plot of $M^2$ for the developed strip is shown at the top of Fig.~\ref{fig:Msq_Lk_1_5}, with the length scale arbitrary, along with three larger widths, including one at self-contact. The symmetry arguments developed above for $\mathrm{Lk}=0.5$ carry over for $\mathrm{Lk}=1.5$ with the replacement $L\rightarrow L/3$. Thus $\kappa, \eta, \eta'$ have the same reflection symmetry as in Fig.~\ref{fig:kappa}, with $L\rightarrow L/3$, and the same symmetry under the translation $s\rightarrow s+L/3$. These induce corresponding parity eigenstates \eqref{chi_refl},\eqref{chi_trans} with $L$ replaced with $L/3$. These are the so-called basically periodic solutions \cite{Arscott}, which allow the domain to be reduced to $L/6$ for numerical integration using the FD method. From the point of view of Bloch's theorem, $\chi(u^1+L/3,u^2)=\exp(ikL/3)\chi(u^1,-u^2)$, with a Bravais lattice vector of magnitude $L/3$. To satsify the periodic boundary condition $kL$ must be an integer multiple of $\pi$. Here we only show solutions for $kL/3=0,\pi$.

\begin{figure}[floatfix]
\includegraphics[width=0.9\textwidth]{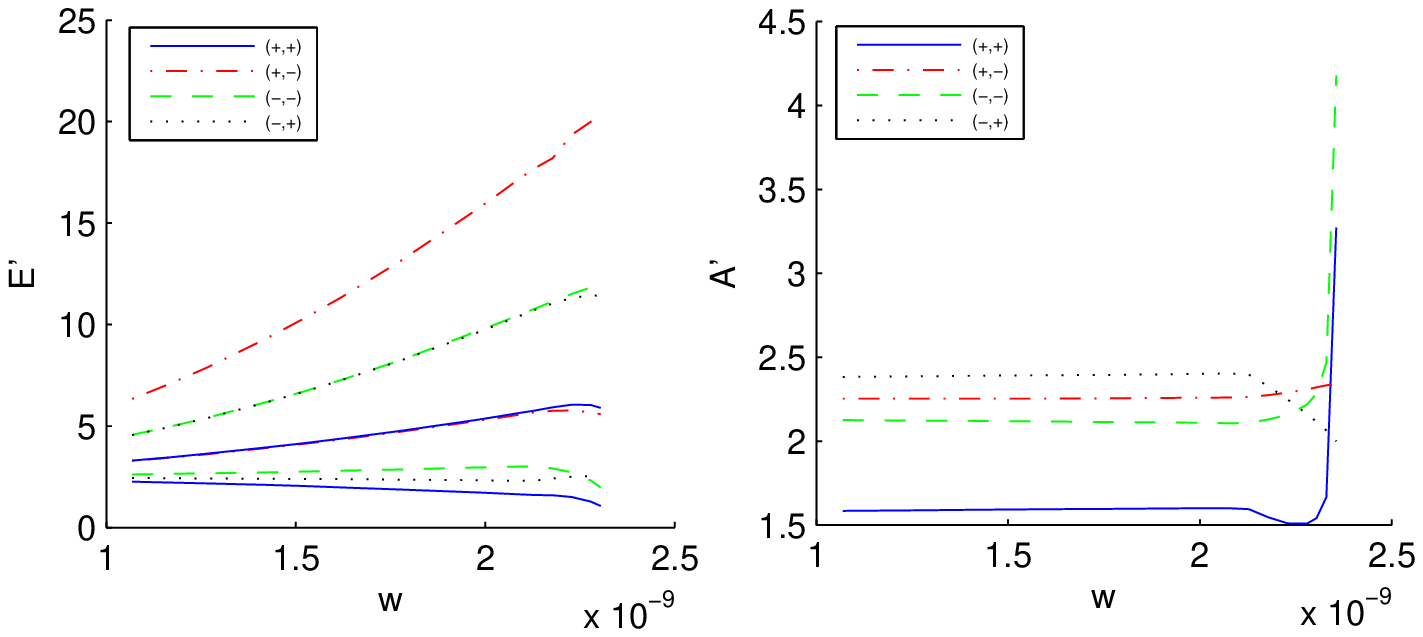}
\caption{\label{fig:AE_Lk_1_5}$E',A'$, $\mathrm{Lk}=1.5$}
\end{figure}

\begin{figure}[floatfix]
$
\begin{array}{c}
\includegraphics[width=0.9\textwidth]{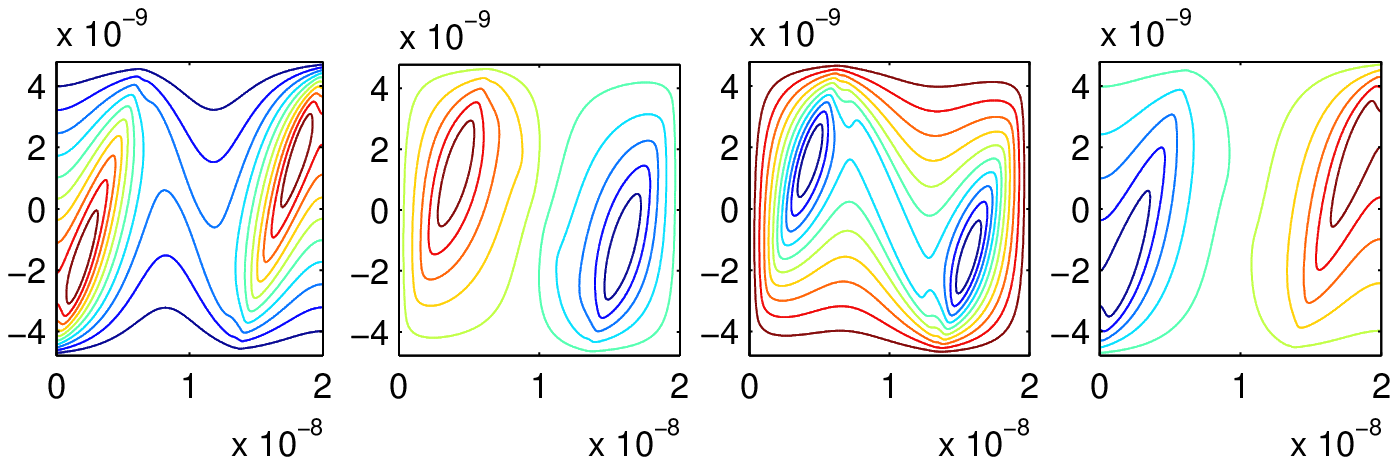}\\
\includegraphics[width=0.9\textwidth]{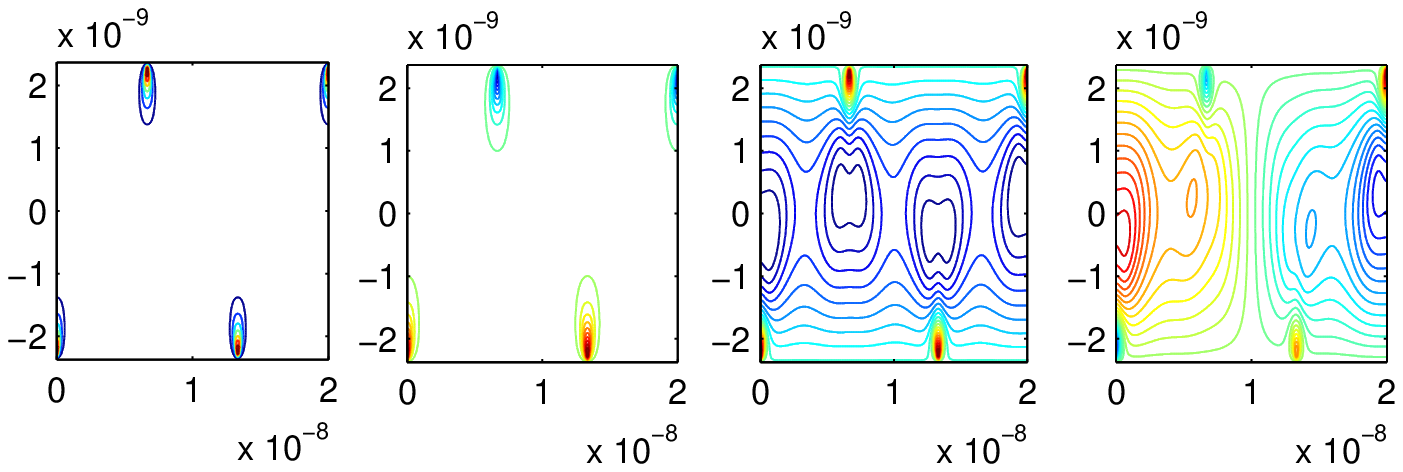}
\end{array}$
\caption{\label{fig:Wfn} $\mathrm{Lk}=0.5$ $(L/2w=2\pi/3),1.5$ $(L/2w=4.2205)$}
\end{figure}

As with Mathieu's equation, solutions of higher period exist \cite{Arscott}, so long as the periodic boundary condition is satisfied. For example, to show solutions symmetric or anti-symmetric under translation $\chi(u^1,u^2)=\pm\chi(u^1+L,-u^2)$ are induced by the underlying symmetry of $\kappa, \eta, \eta'$, note that under $s\rightarrow s+L$, $u^1\rightarrow u^1+L$ and that $M^2$ is invariant ($\eta, \eta'$ and $u^2$ all change sign). Similarly, under the reflection, $s\rightarrow L-s$, $u^1\rightarrow L-u^1$ and $M^2$ is invariant since $\eta$ is even, but $\eta'$ and $u^2$ change sign under translation. From the point of view of Bloch's theorem, this corresponds to a Bravais lattice vector of magnitude $L$. For these higher periods fewer nodes are spread over the same length $L$, so more states of lower energies are observed. Note that period $L/3$ implies period $L$ but not vice versa. Shown in Fig.~\ref{fig:AE_Lk_1_5} (left) is the dimensionless energy $E'$ for $\mathrm{Lk=1.5}$. The large $w$ energy behaviour does not now decrease faster than $w^2$. The (-,-) and (-,+) states also show a degeneracy at large $w$.

\subsection{Curvature trapping of states}
Much attention has been given to the degree of localisation of states of particles constrained to move in structures with curvature \cite{Hurt,Goldstone}, with Costa \cite{Costa} giving the example of one bound state for the bookcover surface with a zero transmission coefficient in the case of an infinitely sharp bend. One measure used for distinguishing between localised and extended states is the inverse participation ratio \cite{Thouless,Taira}, $A=\int |\psi|^4 \mathrm{d}V/(\int |\psi|^2 \mathrm{d}V)^2$. The non-degenerate flat ground state $\chi=\cos(\pi u^2/2w)$ has dimensionless $A'\equiv A(2Lw)=1.5$, whereas the flat higher excited states have $A'=2.25$. 

Shown in Fig.~\ref{fig:AE_Lk_0_5} (right) is $A'$ for $\mathrm{Lk}=0.5$, showing a general trend of increase in localisation for increasing width. Comparison with the flat state values is perhaps not so meaningful for higher excited states as they all have the same inverse participation ratio. The figure shows $A'$, following the lowest energy eigenstate for each parity with increasing width.

Shown in Fig.~\ref{fig:AE_Lk_1_5} (right) is the corresponding $A'\equiv A(2Lw/3)$ for $\mathrm{Lk=1.5}$. The flat values are the same as for $\mathrm{Lk}=0.5$. There is a marked increase of the inverse participation ratio with width, except for the (-,+) state. There is a dip in the localisation of the ground state before it also sharply increases with width.

The corresponding lowest energy wave functions are shown in Fig.~\ref{fig:Wfn} for the highest width M\"obius structures, showing confinement of the wave function to the high curvature regions corresponding to the lowest plot in each of Figs.~\ref{fig:Msq_Lk_0_5}, \ref{fig:Msq_Lk_1_5}. The difference in topology is that $\mathrm{Lk}=1.5$ structures have no creases, prevented by self-contact, giving degenerate, disconnected, wave functions concentrated at the singularities in $M^2$ at the largest width. The $\mathrm{Lk}=0.5$ structures, by contrast, show localisation to the creases formed at higher widths, allowing the wave function to be connected across the whole domain.

\section{Conclusions}
The eigenvalues and eigenstates are calculated for a quantum mechanical particle confined to geometric M\"obius strips of arbitrary width, with $\mathrm{Lk}=0.5, 1.5$, the shape calculated by rigorously minimising the Wunderlich bending energy functional. The eigenstates are found to be localised to the potential wells obtained when increasing the width of the M\"obius structures, leading to increased localisation to the regions of highest curvature. Our geometric formulation could be used to study transport properties of M\"obius strip nanoribbons and molecules used in nanoscale devices \cite{Maiti}.

\bibliography{apsmob2}

\end{document}